\documentclass[prl,twocolumn,showpacs]{revtex4-1}

\usepackage{graphicx}
\usepackage{amsmath}
\usepackage{amssymb}
\usepackage{epsfig}

\renewcommand{\v}[1]{{\bf #1}}

\def\eqa{\begin{eqnarray}}
\def\eea{\end{eqnarray}}
\newcommand{\eq}{\begin{equation}}
\newcommand{\ee}{\end{equation}}

\newcommand{\ua}{\uparrow}
\newcommand{\da}{\downarrow}
\newcommand{\ra}{\rightarrow}

\newcommand{\al}{\alpha}
\newcommand{\bt}{\beta}

\newcommand{\Del}{\Delta}
\newcommand{\eps}{\epsilon}

\newcommand{\ga}{\gamma}

\newcommand{\la}{\lambda}
\newcommand{\La}{\Lambda}

\newcommand{\si}{\sigma}

\usepackage{color}

\begin{document}

\title{Theory of superconductivity in a three-orbital model of Sr$_2$RuO$_4$}
\author{Q. H. Wang${}^{1}$}
\author{C. Platt${}^{2}$}
\author{Y. Yang${}^{1}$}
\author{C. Honerkamp${}^{3,4}$}
\author{F. C. Zhang${}^{5,6}$}
\author{W. Hanke${}^{2}$}
\author{T. M. Rice${}^{5,7}$}
\author{R. Thomale${}^{2,8}$}

\affiliation{${}^{1}$National Laboratory of Solid State Microstructures,
Nanjing University, Nanjing, 210093, China}
\affiliation{${}^{2}$Theoretical Physics, University of W\"urzburg, D-97074
  W\"urzburg, Germany}
\affiliation{${}^{3}$Institute for Theoretical Solid State Physics, RWTH
Aachen University, D-52056 Aachen, Germany}
\affiliation{$^4$JARA - FIT Fundamentals of Future Information Technology, Germany}
\affiliation{${}^{5}$Department of Physics, and Center of Theoretical and
Computational Physics, The University of Hong Kong, Hong Kong,
China}
\affiliation{${}^{6}$Department of Physics, Zhejiang University,
Hangzhou, China}
\affiliation{${}^{7}$Institute for Theoretical Physcis, ETH Zurich, CH-8093  Z\"urich, Switzerland}
\affiliation{${}^{8}$Institut de th\'eorie des
  ph\'enom\`enes physiques, \'Ecole Polytechnique F\'ed\'erale de
  Lausanne, CH-1015 Lausanne, Switzerland}

\date{\today}

\begin{abstract}
In conventional and high transition temperature copper oxide and iron pnictide superconductors, the Cooper pairs all have even parity. As a rare exception, Sr$_2$RuO$_4$ is the first prime candidate for topological chiral $p$-wave superconductivity, which has time-reversal breaking odd-parity Cooper pairs known to exist before only in the neutral superfluid $^3$He. However, there are several key unresolved issues hampering the microscopic description of the unconventional superconductivity. Spin fluctuations at both large and small wavevectors are present in experiments, but how they arise and drive superconductivity is not yet clear. Spontaneous edge current is expected but not observed conclusively. Specific experiments point to highly band- and/or momentum-dependent energy gaps for quasiparticle excitations in the superconducting state. Here, by comprehensive functional renormalization group calculations with all relevant bands, we disentangle the various competing possibilities. In particular we show the small wavevector spin fluctuations, driven by a single two-dimensional band, trigger $p$-wave superconductivity with quasi-nodal energy gaps.
\end{abstract}

%75.30.Fv  Spin-density waves
%74.20.Rp  Pairing symmetries (other than s-wave)
%74.20.-z  Theories and models of superconducting state
%71.27.+a  Strongly correlated electron systems; heavy fermions
%64.60.ae  Renormalization-group theory

\maketitle

Very soon after the discovery of superconductivity in Sr$_2$RuO$_4$~\cite{maeno}, it was proposed that the superconducting (SC) pairing is of unconventional nature~\cite{ricesigrist,baskaran}.  Later experiments have provided evidence that the Cooper pair in the SC state is of odd parity~\cite{liu} with total spin equal to one~\cite{NMR}.  Further evidence indicates the superconductivity to be chiral, breaking time reversal symmetry~\cite{muSR,Kerr}. Sr$_2$RuO$_4$ is thus the first prime candidate for a chiral $p$-wave superconductor~\cite{Mackenzie2003, Bergemann2003, Maeno2012, Kallin2012}, an interesting analogue of the neutral superfluid $^3$He. It has recently received great interest as by suitable manipulations it may support zero energy Majorana bound states in vortices~\cite{ivanov},  the building block for topological quantum computing~\cite{nayak}. However, there are a number of outstanding issues associated with the chiral $p$-wave superconductivity in Sr$_2$RuO$_4$. First, $p$-wave spin triplet pairing is expected to be associated with spin fluctuations at small wavevector. However, the spin density wave (SDW) fluctuation observed in Sr$_2$RuO$_4$ is dominated by a large wavevector at higher temperatures and coexist with a feature at small wavevector at lower temperatures.~\cite{braden} A resolution of this puzzle is vital to understand the superconductivity. Second, one would expect a spontaneous electric current at the edge of the RuO$_2$ layers as a result of the chiral SC state. The edge current, however, has not been observed conclusively in experiments.~\cite{kirtley} One possible reason is the edge current is very fragile and difficult to establish against disorders. Another possibility is a topological cancellation from hole-like and electron-like bands,~\cite{Raghu} posing a question as whether the SC state is topologically nontrivial at all. Third, the specific measurement reveals abundance of low energy quasiparticle excitations below the transition temperature.~\cite{Agterberg97} This would point to multiple gaps of very different magnitudes and/or deep minima in strongly momentum dependent gap functions.
Previous theories treat either the two-dimensional (2D) $\ga$-band derived from the $xy$ orbital,~\cite{Hlubina1999, Nomura2000, Honerkamp2001,nomura2002} or the quasi-one-dimensional (1D) $\al$ and $\bt$ bands derived from the $xz$ and $yz$ orbitals.~\cite{Raghu,huo} However, the evolution of wavevectors of the spin fluctuations is beyond such models, and in fact can only be accounted for by a complete three-band model. This in turn dictates the properties of the SC state mentioned above. {\em A microscopic theory for ruthenate superconductivity should explain both SDW and SC fluctuations at different energy or temperature scales.}
In this paper, we apply the functional renormalization group theory (FRG)~\cite{Honerkamp2001,metzner2012} to study a 3-band Hubbard model including both the 2D-$\ga$ and the quasi-1D ($\al$,$\bt$) Fermi sheets as suggested by first-principle calculations and angle-resolved photoemission (ARPES) experiments.~\cite{Bergemann2003,arpes} The FRG is particularly promising to address the multi-scale energy issues in ruthenates. In recent years the FRG has been expanded to treat 2D multi-orbital systems such as the iron pnictides and selenides ~\cite{fawang,thomale,FeSe-qhwang} and candidate models of topological superconductors with or without time-reversal symmetry~\cite{graphene-thomale,graphene-qhwang,kagome-qhwang,cobalt-thomale,topo-qhwang}. In addition to addressing the pairing symmetry and energy scale, FRG gives information on the relative strength and wavevector of competing orders in the particle-hole channels. Our results show that the SDW fluctuations are driven mainly by two 1D bands at the large wavevector and by the 2D band at the small wavevector, successively as the energy scale is lowered. The latter triggers at an even lower energy scale $p$-wave superconductivity, which is dominated by the 2D band and has a highly anisotropic gap and deep minima near the Brillouin zone boundary.
Our theory predicts chiral edge modes and thus edge current. However, the large gap anisotropy indicates the fragility of the chiral edge modes against perturbations such as disorder, rendering the detection of edge current hard to accomplish. Our prediction on the strong SDW fluctuations at a small wavevector at low temperatures can be tested in further neutron scattering experiment, and the prediction on the strongly anisotropic gap function in momentum space should be tested in ARPES with high resolution at extremely low temperatures.

\begin{figure}
\includegraphics[width=8.5cm]{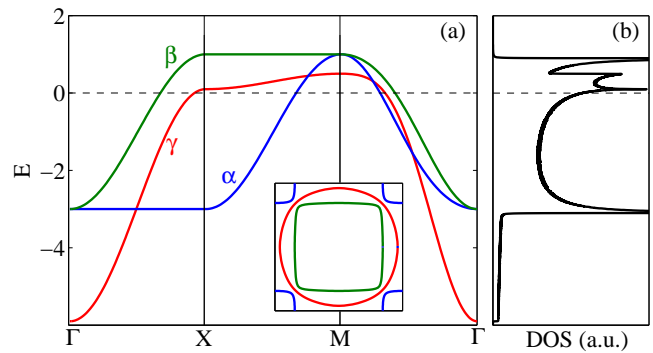}
\caption{Band structure of the Sr$_2$RuO$_4$ model. (a) Dispersion along high
symmetry cuts. (Inset: Fermi surface structure with two quasi 1D bands
and
one 2D band.)
(b) Density of states resulting from (a).}\label{band_sro}
\end{figure}

The model we consider is described by the Hamiltonian
\begin{eqnarray}
H&=&
\sum_{\v k , \si \atop a,b} \psi_\v {k a \si}^\dag \eps^{ab}_\v
k\psi_{\v k b \si}^{\phantom{\dag}}+U\sum_{i,a}n_{ia\ua}n_{ia\da} +
U'\sum_{i,a>b}n_{ia}n_{ib} \nonumber \\
&& +
J\sum_{i,a>b,\si,\si'}
\psi_{ia\si}^\dag
\psi_{ib\si}^{\phantom{\dag}}
\psi_{ib\si'}^\dag
\psi_{ia\si'}^{\phantom{\dag}} \nonumber \\
&& +
J'\sum_{i,a\neq b}
\psi_{ia\ua}^\dag
\psi_{ia\da}^\dag
\psi^{\phantom{\dag}}_{ib\da}
\psi^{\phantom{\dag}}_{ib\ua}.  %+V\sum_{\<ij\>,\sigma, \sigma'} n_{i\sigma} n_{j\sigma'}.
\label{hamiltonian}
\end{eqnarray} Here, $\v k$ denotes the momentum, $\si$ the
spin, $i$ the lattice site, and $a$ and $b$ the orbital labels,
with $\psi_{a=1,2,3}$ annihilating an electron in $d_{xz},\ d_{yz}$ and $d_{xy}$
orbitals, respectively. The local interaction parameters include intraorbital
($U$), interorbital ($U'$), Hund's ($J$), and pair hopping ($J'$).
The matrix dispersion function $\eps^{ab}_\v k$ has the following
nonzero elements: $\eps_\v k^{11}=-2t_1\cos k_x-\mu$, $\eps_\v
k^{22}=-2t_1\cos k_y-\mu$, $\eps_\v k^{12/21}=-4t_2\sin k_x\sin
k_y$, and $\eps_\v k^{33}=-2t_1'(\cos k_x+\cos k_y)-4t_2'\cos
k_x\cos k_y+\Del-\mu$, where in dimensionless units, $t_1=1$,
$t_2=0.1$, $t_1'=0.8$, $t_2'=0.35$, $\Del=-0.2$ is the crystal
field splitting, and $\mu=1.1$ is the chemical potential. This set
of parameters produces the band structure shown in
Fig.~\ref{band_sro}a. The inset shows the Fermi surface, which
resembles closely what is observed
experimentally~\cite{Bergemann2003,arpes}. The corresponding normal
state density of states is shown in Fig.~\ref{band_sro}b. There
are van Hove points at the $X$ points on the $\ga$-band close to
the Fermi level, while the band edge anomalies of the $\al$ and
$\bt$ bands are far from the Fermi level.

As known for such a system with partial nesting
and van Hove singularities near the Fermi level, there will be
various competing and mutually interacting collective fluctuations
in density-wave and pairing channels. This physics can be
investigated appropriately by FRG. It provides coupled flow of wavevector
resolved effective interactions in all particle-particle and particle-hole channels versus a running energy scale $\La$.
We use the singular-mode FRG (SMFRG)~\cite{husemann,FeSe-qhwang,graphene-qhwang} to gain benefit of resolving the interactions
throughout the Brillouine zone in terms of form factors, and use the multi-patch FRG~\cite {Honerkamp2001,metzner2012}
to check, with increased angular resolution, that no important form factors have been left out.
From the combination the dominant ordering tendencies can be most suitably addressed. See Supplementary Materials for technical details.
In both schemes, the effective interaction at a given scale can always
be decomposed as, \eqa V^{ab;cd}(\v k,\v k',\v q)\ra \sum_m
S_m (\v q) \phi_m^{ab}(\v k,\v q)[\phi^{cd}_m(\v k',\v q)]^*, \eea
either in the SC, spin or charge channels. Here $a,b,c,d$ are
orbital or band labels, $\v q$ is the associated collective wavevector, and $\v k$ (or $\v k'$) is an internal momentum of the
fermion bilinears $\psi^\dag_{\v k+\v q,a}\psi^\dag_{-\v k,b}$ and
$\psi^\dag_{\v k+\v q,a}\psi_{\v k,b}$ in the particle-particle
and particle-hole channels, respectively. (The Cooper instability
occurs at $\v q=0$ in the particle-particle channel). The
most attractive or fastest growing one of the eigenvalues $S_m (\v q)$
represents the dominant ordering tendency in the respective
channel. The divergence energy scale is an upper estimate for the
ordering temperature.

\begin{figure}
\includegraphics[width=8.5cm]{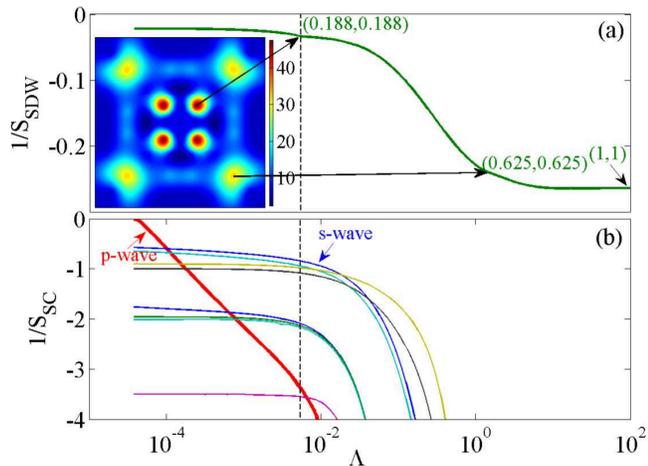}
\caption{(a) FRG flow of the leading eigenvalue $S_{\text{SDW}}(\v q)$ in the
spin channel. The inset shows the $\v q$-map at the instability scale.
(b) Leading pairing eigenvalues $S_{\text{SC}}(\v q=0)$. The thick
line denotes the two eventually diverging $p$-wave pairing modes.
Arrows indicate level crossings associated with the evolution of
the $\v q/\pi$ in the spin channel in (a) and the pairing symmetries
in (b). The vertical dashed line highlights the
correlation between the emergence of small-$\v q$-spin feature and
the $p$-wave pairing tendency.}\label{flow}
\end{figure}

In Fig.~\ref{flow} we show the SMFRG flow of the leading
eigenvalues in the (a) spin and (b) SC channel, for bare
interactions $(U,U',J,J')=(3.2,1.3,0.3,0.3)$. In (a), changes
in the dominant wavevector of the spin interaction are marked by
arrows. At high scales, the spin channel dominates over the SC
channel. The dominant spin-fluctuation wavevector evolves from $\v
q=(1,1)\pi$ to $\v q\sim \v q_1 = (0.625,0.625)\pi$ as $\La$
decreases. For $\La< 5\times 10^{-3}$, a further level crossing to
$\v q\sim \v q_2=(0.188,0.188)\pi$ occurs. We checked the form
factors $\phi_m^{ab}(\v k,\v q)$ to find that the $\v q_2$-feature
comes dominantly from the $\ga$-band, while the $\al$ and $\bt$
bands mainly contribute to the $\v q_1$-feature. The spin response at $\v q_2$ is due to the
proximity to the van Hove singularity in the $\ga$-band mentioned
previously. This spots an effect that can not be detected in an analysis for vanishingly small
interactions \cite{Raghu}, as a finite
interaction scale is needed for the proximate van Hove points to come into
play. The evolution of the spin-fluctuation peak from larger to
small $\v q$ with decreasing energy scale is in good qualitative
agreement with neutron scattering experiments~\cite{braden} where a
similar change is observed as a function of temperature. The
charge channel (not shown) is screened down in the flow and only re-enhanced
weakly as $\La$ decreases. At lowest scales, both spin and charge
channels saturate due to imperfect nesting.

In the inset of Fig.~\ref{flow}, we plot the leading spin-channel
eigenvalues $S_{\rm SDW} (\v q)$ versus $\v q$ at the final stage of the RG
flow. We see that the interaction in the spin channel peaks at $\v
q_2$, but the amplitude at $\v q_1$ is also sizable. In both
cases, the spin bilinears correspond to onsite spins.
The attractive pairing interaction is induced at intermediate
scales via the spin channel. As the dominant spin fluctuation
vector changes during the flow, the dominant pairing fluctuations also
undergo changes as a function of $\La$. In Fig.~\ref{flow}b, we
show the $10$ leading attractive eigenvalues of the pairing
channel. At low scales, the
strongest growing eigenvalue belongs to a $p$-wave
mode which is two-fold degenerate due to the underlying $C_{4v}$ symmetry.
By comparing the flow in the spin channel in
(a), we see that this pairing mode is already seeded and enhanced
as the $\v q_2$-feature shows up (the correlation is shown by the
vertical dashed line), supporting the interpretation that close-to-ferromagnetic spin
fluctuations drive triplet $p$-wave pairing in this case. The final
portion of the SC flow is log-linear in $\La$,
consistent with the fact that the spin and charge channels
saturate and decouple from the SC channel in the lowest energy
range.

\begin{figure}
\includegraphics[width=8.5cm]{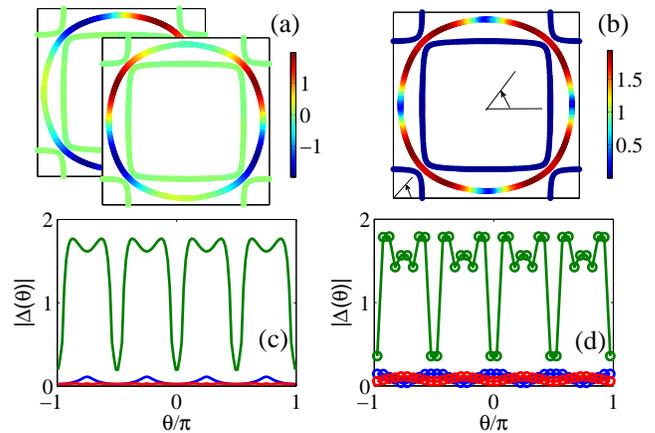}
\caption{The gap structure (not to scale). (a) Leading $p$-wave form factors on the Fermi
surfaces. (b) $p+ip'$ gap amplitude $|\Del(\v k)|$ on the
Fermi surfaces. (c) The Fermi angle $\theta$ dependence of $|\Del(\theta)|$
for the pockets $\al$ (blue), $\bt$ (red) and $\ga$ (green) as
obtained by SMFRG. $\theta$ is defined
according to the arrows in (b). The
amplitudes on $\alpha$ and $\beta$ sheets are enlarged for better visibility.
(d) Same plot as (c) obtained by multi-patch FRG without rescaling of
the $\al$ and $\bt$ sheet.} \label{gap}
\end{figure}

We now analyze the detailed pairing
function of the $p$-wave state. Fig.~\ref{gap}a shows the form factors of the two
degenerate $p$-wave functions. For the Fermiology and interaction regime considered, the gap
function in SMFRG turns out to be much smaller on the $\al$ and $\bt$ bands
than on the $\ga$-band, in general agreement with multipatch FRG.
On the $\ga$ band, the degenerate gap form factors from the SMFRG can be written
approximately as $p_\v k=p_1\sin k_x+p_2\cos k_y\sin k_x$ and
$p'_\v k=p_1\sin k_y+p_2\cos k_x\sin k_y$, where $p_1/p_2=-0.4375$. Thus, it is worth noting that pairing on the next-nearest bond is important.  In the ordered state, as confirmed by a mean field calculation using the
renormalized pairing interaction, the favorable state resulting from
the $p$-wave instability is the chiral $p\pm ip'$ state, as the system
maximizes condensation energy by breaking time-reversal symmetry. The gap amplitude $|\Delta(\v k)|$
in this case is shown in Fig.~\ref{gap}b on the
Fermi surface. For better quantitative clarity, in Fig.~\ref{gap}c
we plot the Fermi angle $\theta$ dependence of $|\Del(\theta)|$
on the Fermi pockets $\al$ (blue), $\bt$ (red) and $\ga$ (green). Since the
amplitudes on the $\al$ and $\bt$ pockets are very small, they are enlarged
(by a factor of $20$) for better visibility.
Near $X/Y$, the $\ga$-band gap amplitude shows
deep minima. This is understood as follows. In general, p-wave pairing is stabilized by attractive
(repulsive) interactions upon forward (backward) scattering. The
umklapp contribution to backward scattering, however, involves only
a small momentum transfer, leading to a destructive interference.
In this sense, the $p$-wave pairing
for such a Fermi surface cannot benefit from the enhanced density
of states near $X/Y$, and correspondingly, the energy scale for it
is small (we get $\sim 0.1$meV). The low critical scale is also consistent with
the late emergence of the small-$q$ spin fluctuations shown in Fig.\ref{flow}a.
The depth of the gap minima is
enhanced by the second nearest-neighbor pairing $p_2$ with
opposite sign to $p_1$. The deep minimum feature is likewise found in
multi-patch FRG (Fig.~\ref{gap}d). There, the angular
variation is found to be slightly stronger than for SMFRG, while the
general behavior is the same. Similarly, also for the multi-patch
FRG, the gap on the $\alpha$ and $\beta$-pockets is rather small, even
below the minimum on the $\gamma$-band. While the qualitative behavior is
similar to that of Ref.~\cite{nomura2002}, the anisotropy and
band-selectiveness of the pairing is even stronger in our
infinite-order approach.

\begin{figure}
\includegraphics[width=8cm]{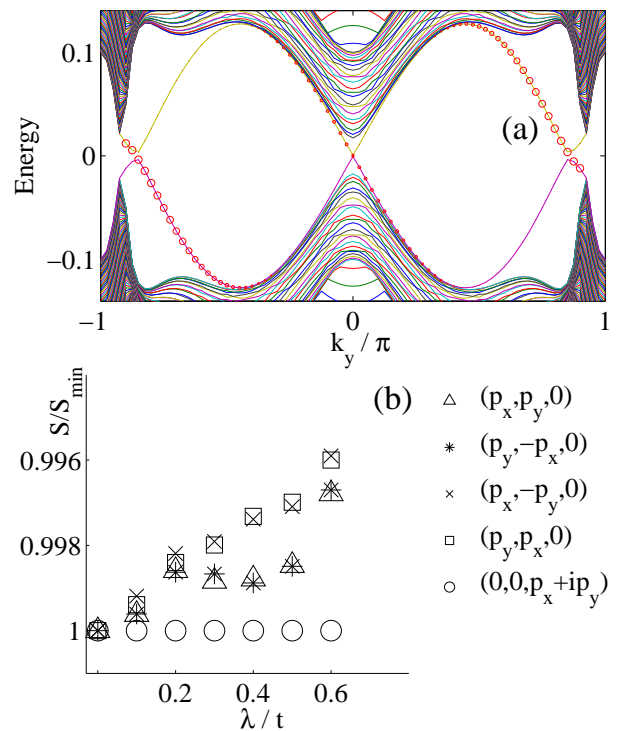}
\caption{(a) Energy spectrum versus the transverse momentum $k_y$ in
the $p+ip'$ SC state of a ribbon under open boundary conditions
along $x$. Only the $d_{xy}$-orbitals are considered. The size of the circles denotes the
wave function amplitude of the low-lying edge states integrated
over two sites nearest to the respective edge. (b) Relative pairing eigenvalues versus
spin-orbit coupling $\la$. The legend shows the five types of $\v d$-vector in the triplet
pairing function. $S_{min}$ is the most attractive eigenvalue.}\label{edge}
\end{figure}

The deep gap minima on the $\ga$-band define a small gap scale of
roughly a tenth of the gap maximum. We discuss two
consequences of this small energy scale: Fig.~\ref{edge} shows the
energy spectrum of the $p+ip'$ SC meanfield Hamiltonian on an
infinite ribbon along the $y$-direction, with open boundary
conditions along $x$. The energy eigenvalues are plotted versus the
transverse momentum $k_y$. The circles denote the amplitude of the
wavefunctions on one of the two edges. We see that there are
subgap edge modes that are unidirectional, i.e. chiral. The
gapless bands localized on either edge cross at $k_y=0$. There are
two additional energy minima of the edge states near $k_y=\pm
\pi$ that appear to be connected to the large second
nearest-neighbor pairing component $p_2$ which in turn enhances
bulk gap minima. Note that the gapless
chiral edge states are protected only up to the deep bulk gap minima, beyond which impurity scattering between the edge modes and the bulk continuum is allowed. The edge modes within the bulk gap minimum may be robust but such low energy Bogoliubov de Gennes quasiparticles are almost charge neutral. This severely reduces the robustness of the chiral edge current with
respect to, {\em e.g.}, edge disorder. Experimentally the detection of the edge current is not yet conclusive.\cite{kirtley}

In thermodynamic quantities, the $\ga$-band contributes
significantly once the temperature is about the SC gap scale,
in addition to the $\al$-and $\beta$-bands. In particular, while there are still
open questions about the role of $(\al,\bt)$ bands, the deep gap minima might contribute
to explaining the power-law behavior in the specific heat at temperatures above the
small gap scale.~\cite{nomura2002} This argument is similar in spirit to a recent discussion of
possible anisotropic chiral $d$-wave superconductivity in sodium cobaltates~\cite{cobalt-thomale}.

Before closing we emphasize that the results for these qualitative
features of the gap structure discussed so far are quite generic. (A
detailed discussion of the general phase diagram of the given
Fermiology beyond the specific ruthenate setting will be given elsewhere.)
We obtain rather similar results for a considerable range of interactions, {\em e.g.},
$(U,U',J,J')=(3.3,1.1\pm 0.1,0.115,0.115)$, where the only
varying aspect we find from the data is the absolute instability scale
of superconductivity. A general trend we find is that the Hund's rule
coupling $J$ and pair hopping $J'$ favor large-$\v q$ SDW interactions, and if
sufficiently large, would destabilize the $p$-wave pairing. Atomic spin-orbital
coupling and inter-layer hopping mixes $d_{xz}/d_{yz}$ ($n=1,2$)
and $d_{xy}$ ($n=3$) orbitals at the one-particle level, leading to inter-band proximity effect between the active and passive
bands which may contribute to the pairing amplitude on the $\al$-and
$\beta$-bands~\cite{Zhit2001}. (At the level of interactions, the pair
hopping $J' $ is the only coupling which would allow for a proximity effect
induced by the $\ga$-band, and the effect is weak since this
coupling is initially orthogonal to the $p$-wave channel.) To
a good approximation, such effects can be included by using the above renormalized pairing
interaction at a suitable energy scale $\La_0$, and continue the
flow in the pairing channel alone, since the particle-hole channel
is essentially saturated and decoupled from the pairing channel at
this stage. The behavior of the eigenvalues of the pairing channel as a function of
spin-orbit coupling strength
$\frac{i\lambda}{2}\epsilon_{abn} \sigma^n_{ss'} \psi^\dagger_{\v
k,a,s} \psi_{\v k,b,s'}$ (where $\eps$ is the antisymmetric tensor
and $\si$ is the Pauli matrix, and repeated orbital and spin
labels are implicitly summed over) is shown in Fig.~\ref{edge}b.
This leads us to the conclusion that the most favorable triplet
pairing $\v d$-vector is $\v d_\v k =(p_\v k\pm ip'_\v k)\hat{z}$,
as already found in previous works based on qualitative arguments. However,
given the small splitting in the eigenvalues, an applied magnetic field could
weaken the effect of spin-orbital coupling in favor of Majorana zero modes in vortices~\cite{ivanov,nayak}.

To conclude, we studied the pairing mechanism in Sr$_2$RuO$_4$ within a
three-orbital model by extensive FRG calculations that go beyond
previous approaches for this system in that they take into account
all three relevant bands and the competition between various
interaction channels to arbitrary order in the bare couplings. The
different FRG approaches we employ show the same trends in the dominant $p$-wave
pairing and SDW channels: We find that the momentum $\v Q$ of
the dominant SDW interaction evolves from $\v
Q\sim (2/3,2/3)\pi$ at high energy scales to $\v Q\sim
(1/5,1/5)\pi$ at low scales. The small-$\v Q$ SDW fluctuations
drive $p$-wave Cooper pairing predominantly on the
$\ga$-band derived from the $d_{xy}$-orbital. The
pairing receives contributions from first and second-nearest neighbors on the Ru square lattice. The
energetically most favorable combination of a $p\pm ip'$-gap
function has deep minima in amplitude on the $\ga$-Fermi surface
near $(\pi,0)$ and $(0,\pi)$. This makes the chiral edge modes
fragile already against a moderate amount of impurities. \\

\acknowledgments{We thank S. A. Kivelson, J. X. Li and M. Sigrist for interesting discussions. All authors agree to the contents of the paper, and have contributed to the paper extensively. TMR, FCZ and QHW initiated the project by using SMFRG, and WH,  CH and RT initiated the project by using patch fRG.  QHW and YY performed SMFRG calculations, and RT and CP performed patch fRG calculations. TMR and CH contributed to the coordination of SMFRG and fRG calculations.  QHW, TMR, FCZ, CH, and RT contributed to the writing.  There are no financial conflicts among the authors. All correspondence should be addressed to TMR (rice@phys.ethz.ch) and/or QHW (qhwang@nju.edu.cn). This work was supported by NSFC (under grant No.10974086, No. 11274269 and No.11023002), the Ministry of Science and
Technology of China (under grant No.2011CBA00108 and 2011CB922101), DFG FOR 723, 912 and SPP 1458, and by the Swiss
Nationalfonds.}

\end{document}